\documentclass[prl,letterpaper,twocolumn]{revtex4-1}
\usepackage{color}
\usepackage{amsmath}
\usepackage{graphicx}
\usepackage{dcolumn}
\usepackage{bm}
\usepackage{subfigure}

\newcommand{\abs}[1]{\left| #1 \right|}

\newcommand{\bv}[1]{\mathbf{#1}}

\newcommand{\bvk}{\bv{k}}

\begin{document}
\title{\bf Dirac Semimetals in Two Dimensions}
\author{Steve M. Young}
\affiliation{Center for Computational Materials Science, U.S. Naval Research Laboratory, Washington, D.C. 20375, USA}
\author{Charles L. Kane}
\affiliation{Department of Physics and Astronomy, University of Pennsylvania, Philadelphia, Pennsylvania 19104-6323, USA}
\begin{abstract}
Graphene is famous for being a host of 2D Dirac fermions.  However, spin-orbit
coupling introduces a small gap, so that graphene is formally a quantum spin Hall
insulator.  Here we present symmetry-protected 2D Dirac semimetals, which feature Dirac cones at high-symmetry points that are \emph{not} gapped by spin-orbit interactions and exhibit behavior distinct from both graphene and 3D Dirac semimetals.  Using a two-site tight-binding model, we construct representatives of three possible distinct Dirac semimetal phases, and show that single symmetry-protected Dirac points are impossible in two dimensions. 
An essential role is played by the presence of nonsymmorphic space group symmetries.  We argue that these symmetries tune the system to the boundary between a 2D topological and trivial insulator.  By breaking the symmetries we are able to access trivial and topological insulators as well as Weyl semimetal phases.
\end{abstract}
\maketitle

Over the past decade, graphene has attracted intense interest as
a material with Dirac cones at the Fermi energy and, as a consequence, a number of unique electronic properties~\cite{Novoselov_2005,Castro_Neto_2009}.      
The Dirac points in graphene, as in similar materials~\cite{Liu_2011,Malko_2012,Zhou_2014,Liu_2015}, are protected by symmetry, but only in the absence of spin-orbit coupling.  Spin-orbit coupling opens a gap at the Dirac point, leading to a topological insulating phase~\cite{Kane05p146802,Kane05p226801}.  The discovery of topological insulators (TIs) heightened interest in three-dimensional Dirac semimetals, which host 3D Dirac points when spin-orbit coupling is included~\cite{Young_2011,Young_2012}. The concept of Dirac and Weyl superconductors has also been recently introduced~\cite{Yang_2014}.
In this Letter we introduce a system that has symmetry-protected 2D Dirac points in the presence of spin-orbit coupling and provide a classification of such systems in general.  These are of interest because they are symmetry tuned to the boundary between topological and trivial insulating phases.

Three-dimensional Dirac semimetals fall into two distinct classes.   In Ref.~\cite{Young_2012} we introduced a Dirac semimetal with Dirac points at high-symmetry points on the surface of the Brillouin zone (BZ).  Candidate materials include $\beta$-cristobalite BiO$_2$~\cite{Young_2012}, as well as distorted spinel materials such as  BiZnSiO$_4$~\cite{Steinberg_2014}.    In these materials the semimetallic state is at the boundary between strong and weak topological insulating phases, and an essential role is played by the nonsymmorphic symmetry of the crystal space group.    A distinct class of Dirac semimetals was introduced in Refs.~\cite{Wang_2012,Wang_2013}, and has been observed in Cd$_3$As$_2$  and Na$_3$Bi~\cite{Liu_2014_1,Liu_2014_2,Borisenko_2014}.  Here, the Dirac points arise due to a band inversion and occur at a generic point on a $C_3$ symmetry axis in the interior of the BZ.  Opening a gap by lowering the symmetry in these materials necessarily leads to a topological insulator -- the trivial insulator is not adjacent.  The 2D Dirac semimetals we introduce here are analogous to the former class:  they arise due to a nonsymmorphic symmetry that requires the conduction and valence bands to touch and exist in the presence of significant spin-orbit coupling.  
We will argue that the nonsymmorphic character correlates with the fact that they are at the boundary between a trivial and topological insulator.    
We will begin by clarifying the role of nonsymmorphic symmetries in protecting Dirac points.   We will then introduce a simple model system that allows us to characterize the allowed Dirac phases in 2D, and conclude with a brief discussion of the possible material venues for these phases, including the layered iridium oxide superlattices recently proposed and studied in Ref.~\cite{kee}.
\begin{figure}
\subfigure[]{\includegraphics[width=4cm]{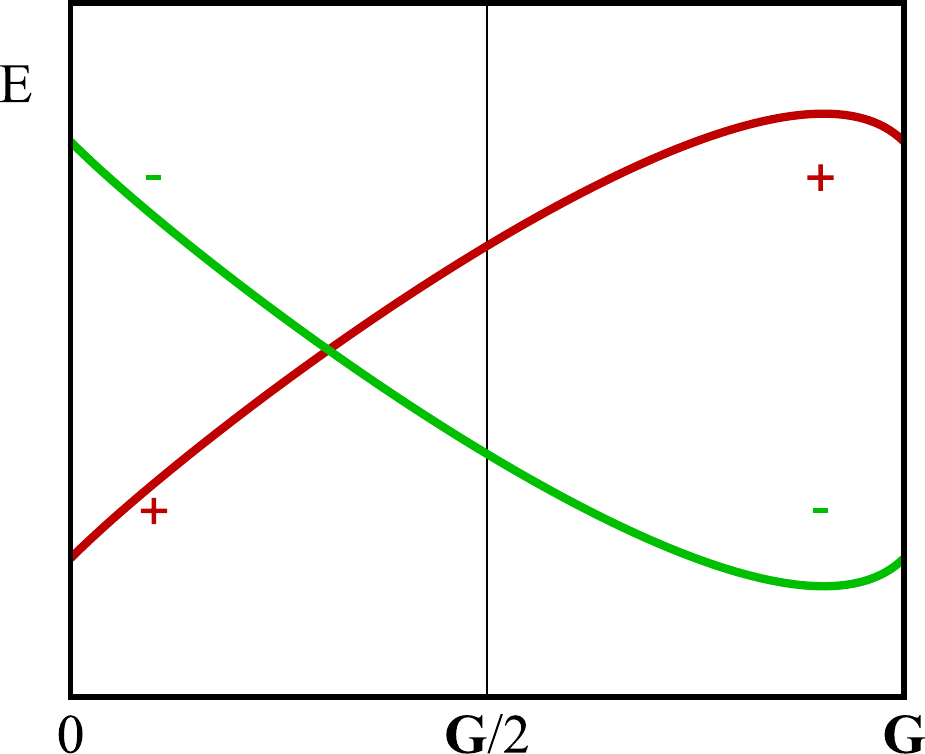}\label{fig:cb_a}}
\subfigure[]{\includegraphics[width=4cm]{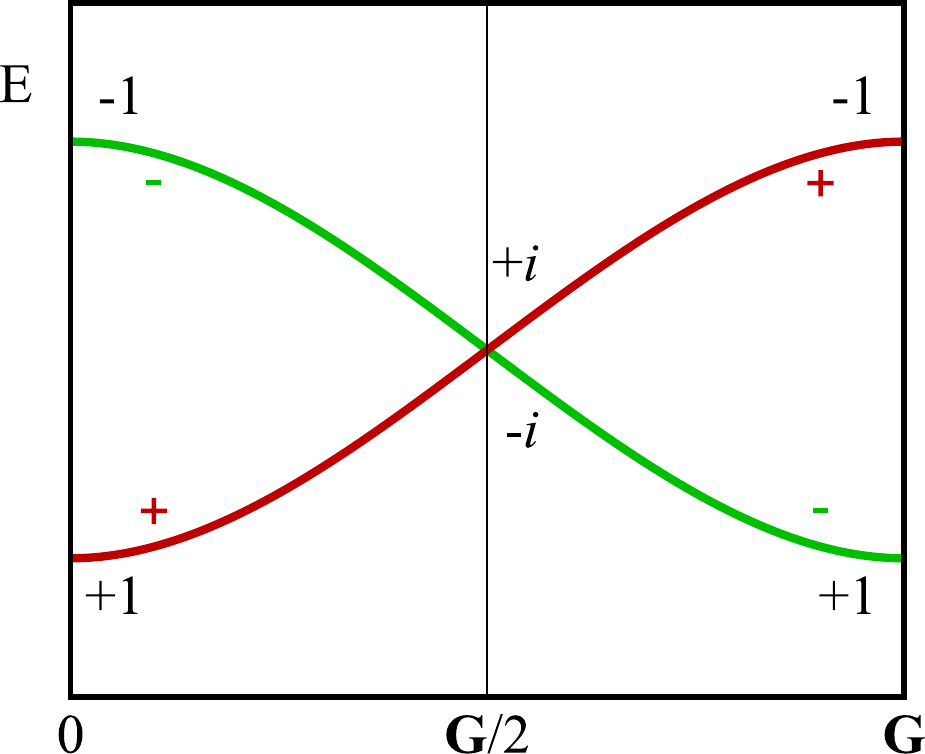}\label{fig:cb_b}}
\subfigure[]{\includegraphics[width=4cm]{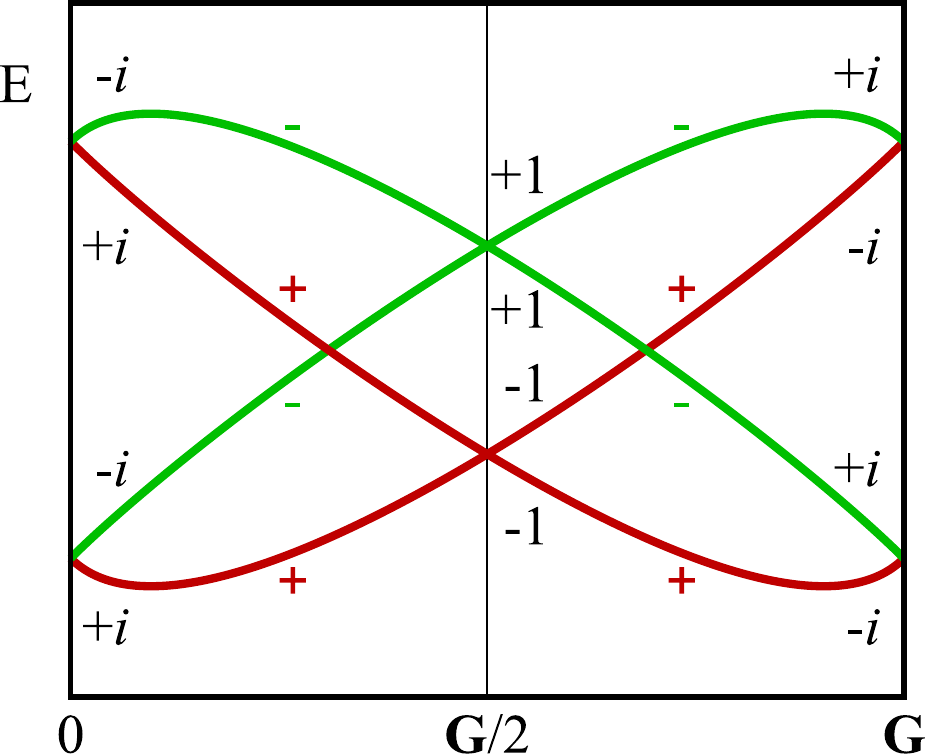}\label{fig:cb_c}}
\subfigure[]{\includegraphics[width=4cm]{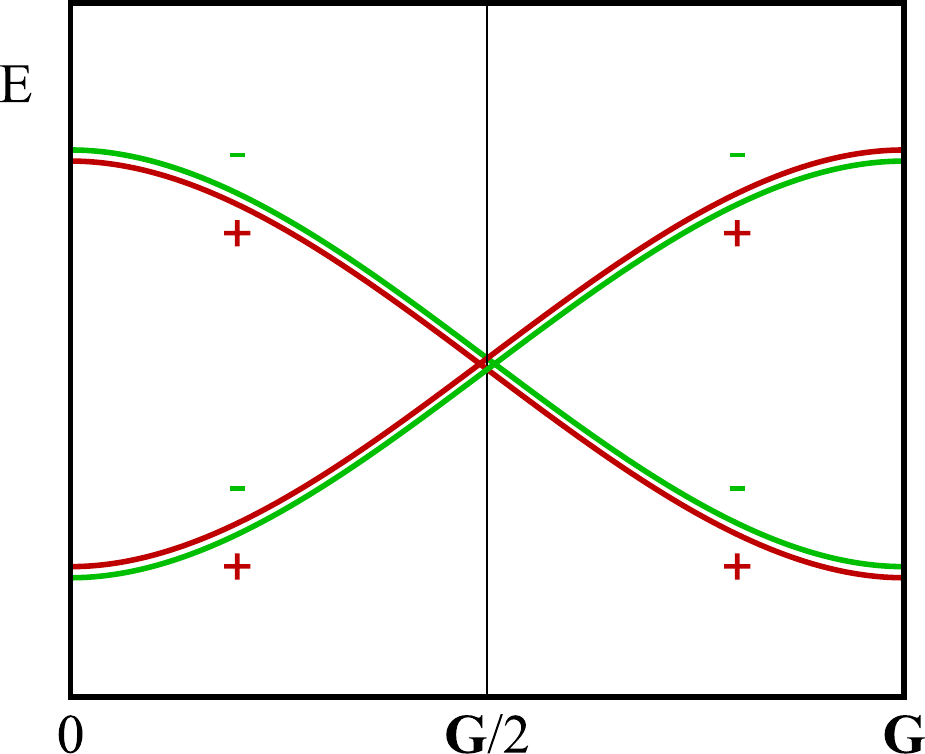}\label{fig:cb_d}}

\caption{A nonsymmorphic symmetry $\{g|{\bf t}\}$ leads to band crossings on a $g$ invariant
line or plane in momentum space.   \subref{fig:cb_a} Without other symmetries, pairs of bands intersect an odd number of times as they cross the BZ.   \subref{fig:cb_b} With time-reversal symmetry, with $\Theta^2=+1$, the crossing occurs at the zone boundary ${\bf G}/2$, where $e^{i{\bf G}\cdot{\bf t}} = -1$.   For $\Theta^2 = -1$, Kramers pairs at ${\bf k}=0$ and ${\bf G}/2$ connect as in ~\subref{fig:cb_c}, leading to a line node in an invariant plane or a Weyl node on an invariant line.   The labels indicate the eigenvalues $\pm \lambda e^{i{\bf k}\cdot{\bf t}}$ of $\{g|{\bf t}\}$.  \subref{fig:cb_d}  With inversion and time-reversal symmetry all states are degenerate (they are offset for clarity), and the crossing occurs at ${\bf G}/2$.}

\end{figure}

It has long been known that nonsymmorphic symmetries lead to extra degeneracies in electronic band structures that cause bands to ``stick together'' due to the existence of higher-dimensional projective representations of the little groups of certain values of ${\bf k}$\cite{Bradley72p1}.  This fact can be understood as a simple consequence of fractional translation symmetries.  Nonsymmorphic space groups are distinguished by the existence of symmetry operations that combine point group operations $g$ with translations ${\bf t}$ that are a fraction of a Bravais lattice vector.   In 2D, the relevant operations, denoted $\{g|{\bf t}\}$, are screw axes $g = C_{2\hat n_\perp}$ ($\hat n_\perp \perp \hat z$), glide mirror lines $g = M_{\hat n_\perp}$ and glide mirror planes $g = M_{\hat z}$, in conjunction with a half-translation ${\bf t}$ that satisfies $g {\bf t} = {\bf t}$ as well as $e^{ i{\bf G}\cdot {\bf t}} = -1$ for the ``odd" reciprocal lattice vectors ${\bf G}$.  The consequence of the fractional translation symmetry is simplest in the case where the unit cell of simple symmorphic crystal is doubled.  In that case, the folded back bands necessarily touch on lines in momentum space.   If the symmetry is reduced so that $\{E|{\bf t}\}$ is violated, the degeneracy is, in general, split. However a nonsymmorphic symmetry $\{g|{\bf t}\}$ still protects degeneracies in the invariant line or plane in the BZ that satisfies $g {\bf k} = {\bf k}$.   In this invariant space the Bloch states can be chosen to be eigenstates $\{g|{\bf t}\} |u^\pm_{\bf k}\rangle = \pm \lambda e^{i{\bf k}\cdot{\bf t}} |u^\pm_{\bf k}\rangle$.  For $\bvk \rightarrow \bvk+{\bf G} $, with $e^{i{\bf G}\cdot{\bf t}} = -1$, the two eigenstates must switch places\cite{ashvin}.  In the absence of other degeneracies pairs of band branches must intersect an odd number of times as they cross the BZ, as shown in Fig.~\ref{fig:cb_a} .

Time-reversal symmetry $\Theta$ imposes further constraints.   The situation is simplest in the absence of spin-orbit interactions, where effectively $\Theta^2 = +1$, and $g^2 = 1$, so $\lambda = \pm 1$.   In this case the degeneracy must occur at the time-reversal invariant momentum ${\bf k}={\bf G}/2$, at the BZ boundary~[Fig~\ref{fig:cb_b}].  At that point the eigenstates of $\{g|{\bf t}\}$ with eigenvalue $\pm i$ are interchanged by $\Theta$.   Stated another way, at ${\bf k}={\bf G}/2$, the Bloch Hamiltonian commutes with $\tilde \Theta = \{g|{\bf t}\} \Theta$, which satisfies $\tilde\Theta^2 = -1$, guaranteeing a Kramers degeneracy.

Spin-orbit interactions lead to additional splitting of the bands, though time-reversal symmetry (with $\Theta^2 = -1$) enforces Kramers degeneracies at the time-reversal invariant momenta.   Moreover, a mirror or twofold rotation satisfies $g^2=-1$, so that $\{g|{\bf t}\}$ has eigenvalues $\pm i$ at ${\bf k}=0$, and eigenvalues $\pm 1$ at ${\bf k}={\bf G}/2$.   It follows that Kramers partners at $\bvk=0$ have \emph{opposite} eigenvalues under $\{g|{\bf t}\}$, while Kramers partners at $\bvk = \pm {\bf G}/2$ have the \emph{same} eigenvalue under $\{g|{\bf t}\}$.    In the absence of other symmetries, this leads to the pattern of splitting shown in Fig.~\ref{fig:cb_c}.  This results in the sticking of four bands, so that a system with the nonsymmorphic symmetry and time-reversal that has a band filling of $4n+2$ (for integer $n$) is necessarily a topological semimetal.   Note that for a glide mirror plane $g = M_{\hat z}$ this leads to a {\it line} of degeneracies, while for a glide mirror line $g = M_{\hat n_\perp}$, it leads to (Weyl) point degeneracies.

If in addition the crystal has inversion symmetry $P$, then since $(P\Theta)^2 = -1$, the bands are Kramers degenerate for all ${\bf k}$.   This leads to a fourfold degenerate crossing at ${\bf k} = {\bf G}/2$~[Fig.~\ref{fig:cb_d}].   
At ${\bf k}={\bf G}/2$,  $P$ and $\{g|{\bf t}\}$ both have eigenvalues $\pm 1$ and commute with $\Theta$.   In addition they anticommute with each other, which guarantees a fourfold degeneracy.
Since the fourfold degeneracy will be split away from ${\bf G}/2$, this constitutes a 2D Dirac point.

\begin{figure}
\subfigure[]{\includegraphics[width=3.5cm]{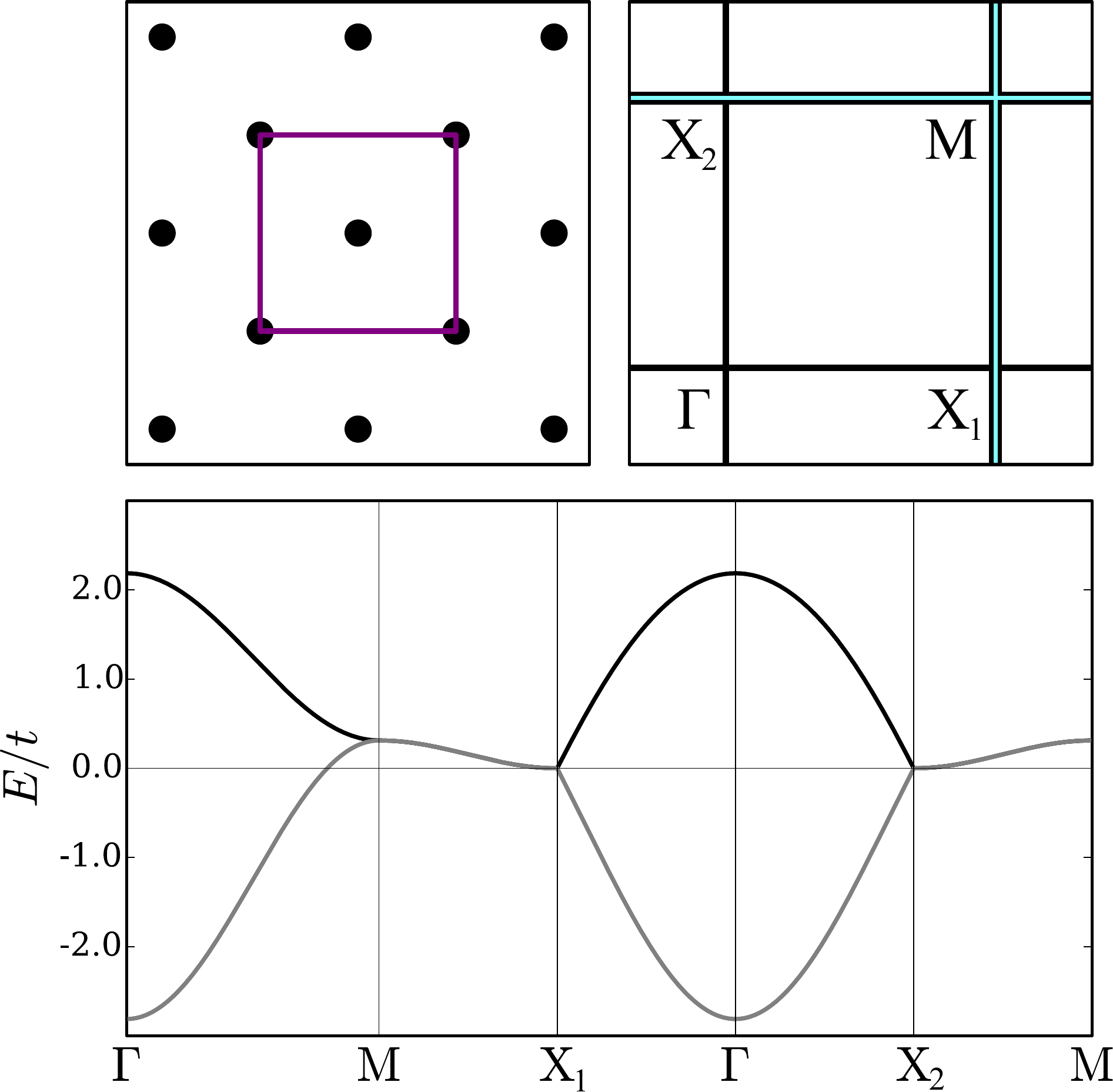}\label{fig:bands_ns}}
\subfigure[]{\includegraphics[width=3.5cm]{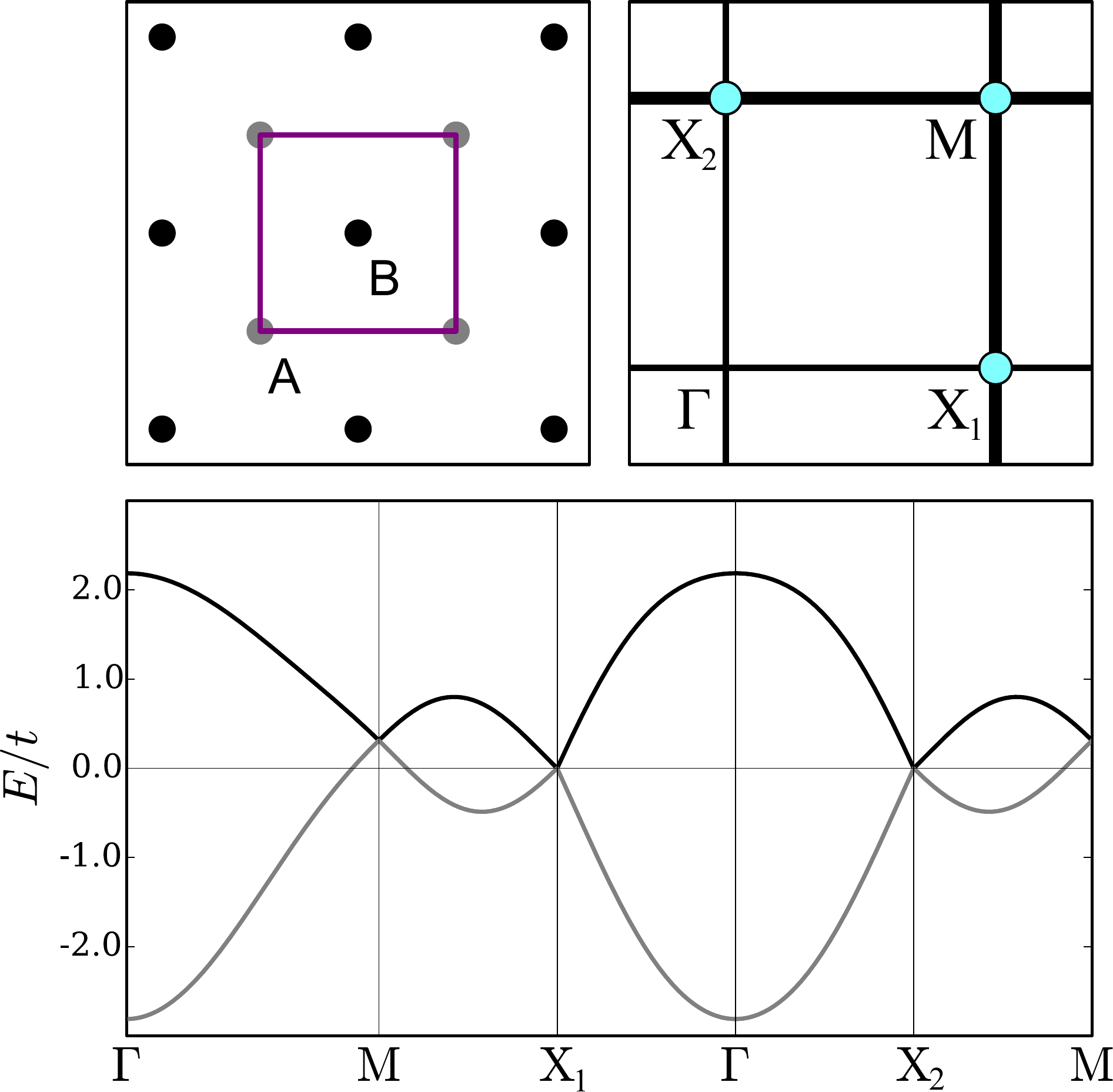}\label{fig:bands_0}}
\subfigure[]{\includegraphics[width=3.5cm]{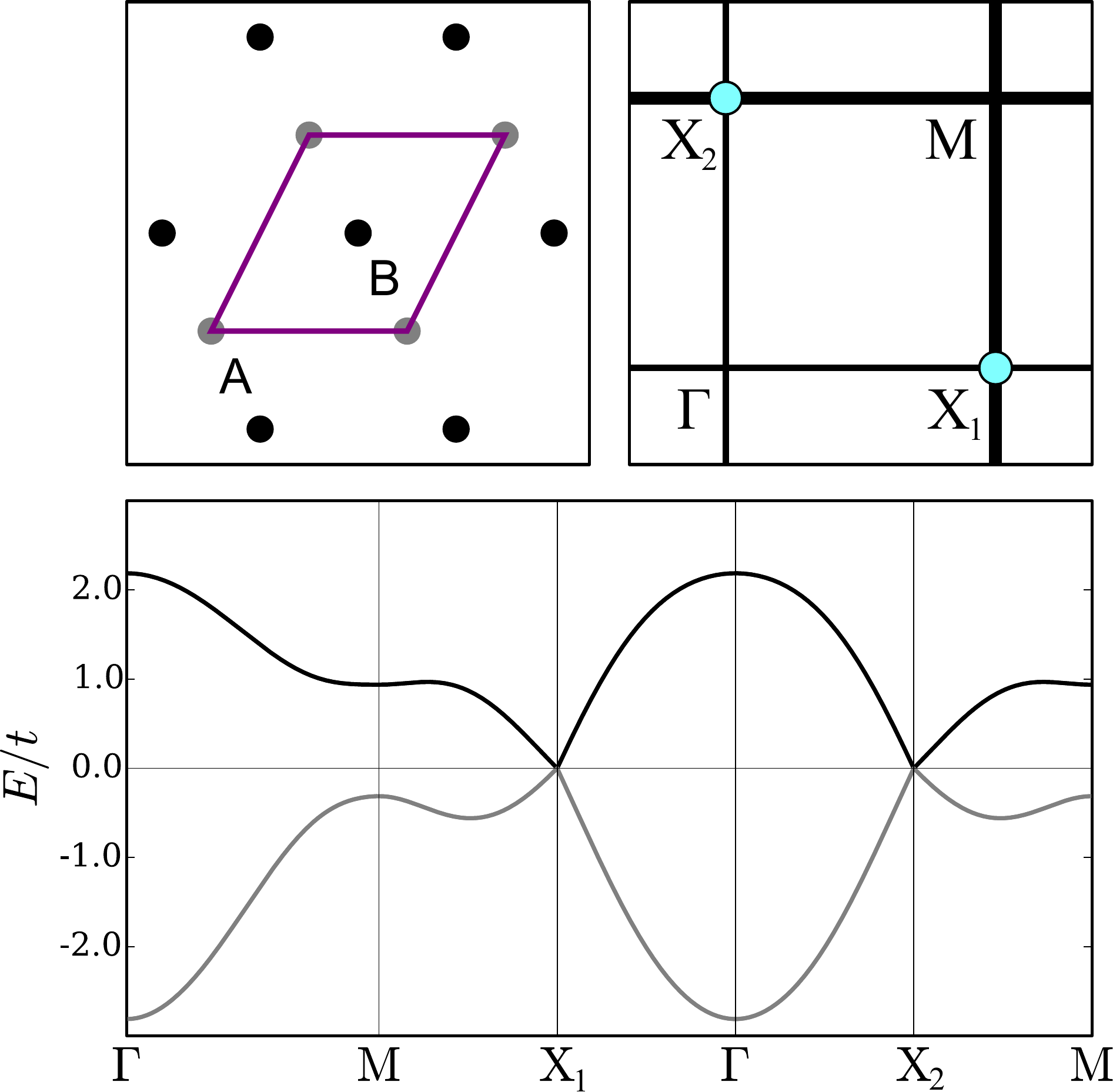}\label{fig:bands_1}}
\subfigure[]{\includegraphics[width=3.5cm]{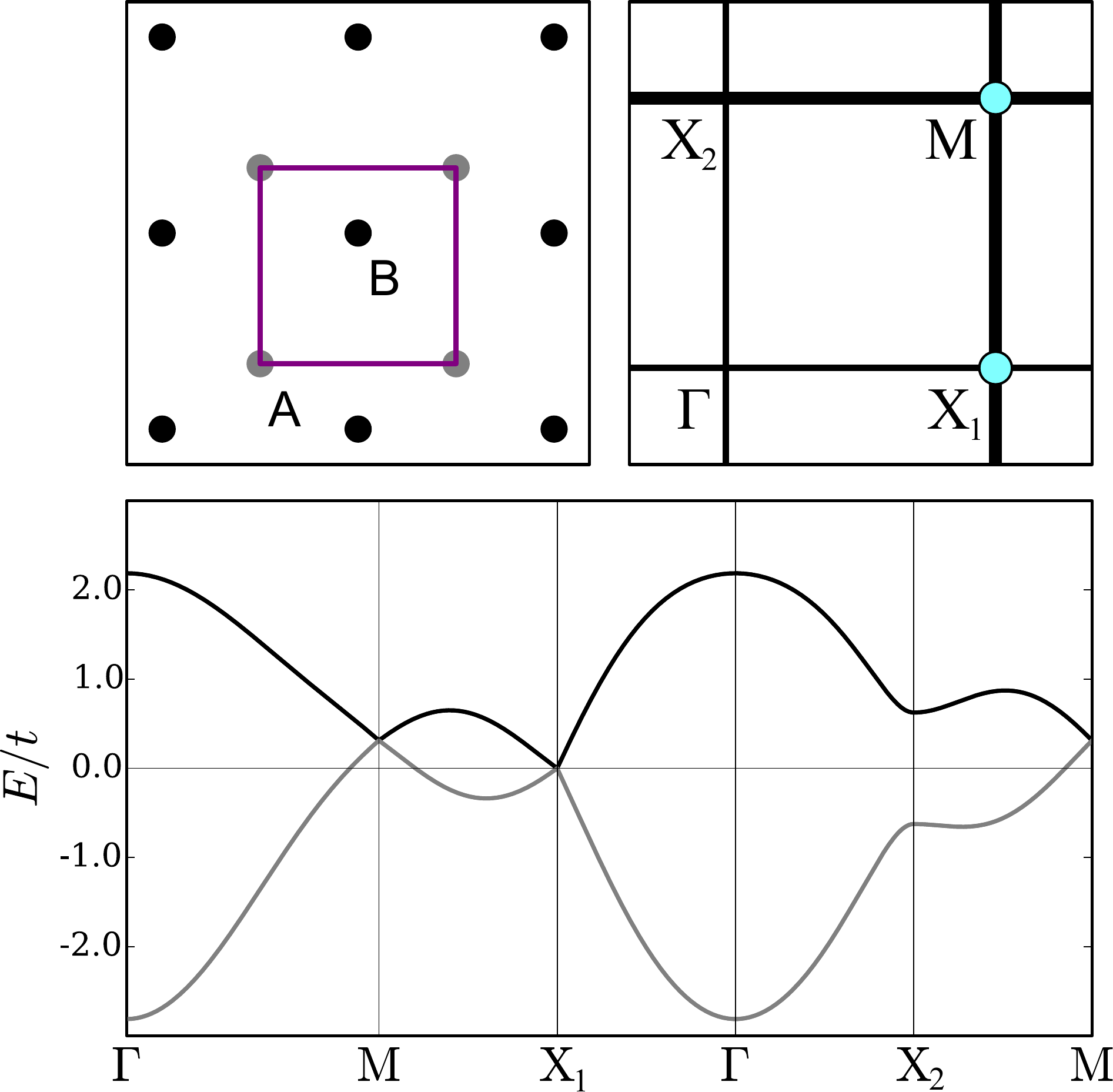}\label{fig:bands_2}}

\caption{ Energy bands for structures with Dirac points protected by inversion symmetry, with lattice structure and BZ shown above.  Dirac points and nodal lines in the BZ are marked in cyan. ~\subref{fig:bands_ns} The $\sqrt{2}\times\sqrt{2}$ supercell of a square lattice with nodal lines along the BZ edge protected by $\{E|{\bf t}\}$.~\subref{fig:bands_0} Crinkling the lattice breaks $\{E|{\bf t}\}$, leaving nonsymmorphic symmetries and Dirac points at $X_1$, $X_2$, and $M$.~\subref{fig:bands_1}  Distorting in the $\langle 11 \rangle$ directions eliminates screw axes $\{C_{2x}|\frac{1}{2}0\}$ and $\{C_{2y}|0\frac{1}{2}\}$ (as well as  $C_4$), gapping the corner Dirac point.~\subref{fig:bands_2}  Alternatively, displacing one of the sites along $\langle 10 \rangle$ breaks $\{M_{\hat z}|\frac{1}{2}\frac{1}{2}\}$, leaving Dirac points at the corner and one edge.}\label{fig:cartoon}
\end{figure}

We now introduce a simple tight-binding model for a 2D Dirac semimetal.  
This can be viewed as a 2D analog of the diamond lattice model for a 3D Dirac semimetal\cite{fukanemele}.   We begin with a square lattice of $s$ states with first and second neighbor hopping and create a $\sqrt{2}\times\sqrt{2}$ unit cell~[Fig.~\ref{fig:bands_ns}].
This features line nodes along the BZ boundary.  
To make the lattice nonsymmorphic we displace the $A (B)$ sublattices in the $+\hat{z} (-\hat{z})$ direction.  This crinkling of the lattice~[Fig.~\ref{fig:bands_0}] permits a second neighbor spin-orbit interaction\cite{Kane05p226801}, leading to the Hamiltonian
\begin{flalign}
H =& 
2t\tau_x\cos\frac{k_x}{2}\cos\frac{k_y}{2} 
+t_2 (\cos k_x + \cos k_y)\nonumber
\\
&+ t^{\rm SO}\tau_z[\sigma_y\sin k_x-\sigma_x\sin k_y ],
\label{h0}\end{flalign}
where $\tau$ and $\sigma$ are Pauli matrices describing the lattice and spin degrees of freedom.
As shown in Fig.~\ref{fig:bands_0}, nonzero $t^{\rm SO}$ breaks the degeneracy on the zone boundary everywhere except the corners $M$ and edge midpoints $X_1$ and $X_2$, at which appear Dirac points.

The present lattice has high symmetry (layer group $P4/nmm$) with multiple symmetries protecting the Dirac points.   From the analysis in Fig.~\ref{fig:cb_d}, the combination of $P$, $\Theta$, and the glide mirror plane symmetry $\{M_{\hat z}|\frac{1}{2}\frac{1}{2}\}$ (${\bf t}$ is in units of the Bravais lattice constant) 
protects the Dirac points at $X_1$ and $X_2$, while the screw axes $\{C_{2\hat x}|\frac{1}{2} 0\}$ and $\{C_{2\hat y}|0\frac{1}{2}\}$ protect the Dirac points at $X_1, M$ and $X_2, M$, respectively.
This can be further seen by examining the $k\cdot p$ theory near these points.  Near ${\bf k}=M$,
\begin{equation}
{\cal H}(M+{\bf q}) = t^{\rm SO} ( \tau_z\sigma_y q_x - \tau_z\sigma_x q_y ).
\end{equation}
At $M$, the symmetries $\Theta = i\sigma_y K$ and $P = \tau_x$ allow a single mass term $\tau_x$.   
This is forbidden by 
$\{C_{2\hat x}|\frac{1}{2} 0\} =  \tau_y \sigma_x $ and 
$\{C_{2\hat y}|0\frac{1}{2}\} =  \tau_y \sigma_y $,
but is allowed by
$\{M_{\hat z}|\frac{1}{2}\frac{1}{2}\} = i\tau_x\sigma_z$.
Likewise, near ${\bf k} = X_1$,
\begin{equation}
{\cal H}(X_1 + {\bf q}) = (t \tau_x - t^{\rm SO} \tau_z \sigma_y) q_x + t^{\rm SO} \tau_z \sigma_x q_y.
\end{equation}
At $X_1$, $\Theta = i\tau_z\sigma_y  K$ and $P = \tau_y$ allows the mass terms $\tau_y$.   This is forbidden by
$\{C_{2\hat x}|\frac{1}{2} 0\} =  \tau_x \sigma_x $ and $\{M_{\hat z}|\frac{1}{2}\frac{1}{2}\} = \tau_x\sigma_z$, but is allowed by $\{C_{2\hat y}|0\frac{1}{2}\} = i \tau_y \sigma_y $.  A similar analysis applies to $X_2$.

In addition to spatial symmetries, Eq.~\eqref{h0} exhibits a particle-hole symmetry when $t_2=0$, manifested by $\{{\cal H},\tau_y\}=0$.   This guarantees the Dirac points at $X_1$, $X_2$, and $M$ occur at the same energy.  $t_2$ violates this symmetry and leads to a shift in the energy at $M$ relative to $X_{1,2}$.    However, the mirror lines $M_{\hat x\pm\hat y}$ guarantee the equivalence of $X_{1,2}$.  Thus, while the touching of the conduction and valence band at $X_1$, $X_2$, and $M$ is guaranteed, there will in general be electron and hole pockets with a finite Fermi surface ~[Figs.~\ref{fig:bands_0} and~\ref{fig:bands_2}].   Nonetheless, with appropriate band structure engineering it may be possible to tune the edge and corner Dirac points to approximately the same energy.   In the following we will explore the range of behaviors that can arise when the symmetries in Eq.~\eqref{h0} are systematically lowered.   We show that Eq.~\eqref{h0} lies at the boundary of three distinct Dirac semimetal phases, two with a pair of Dirac points and the third with three, and prove that in 2D there cannot exist a single symmetry-protected Dirac point.

{\it Case I: Two symmetry equivalent Dirac points}.--First, we consider a distortion that breaks the symmetry between interactions in the $\langle 11\rangle$ and $\langle\bar{1}1\rangle$ directions~[Fig.~\ref{fig:bands_1}], but preserves the mirror line $M_{\hat x+\hat y}$.   
This allows a distortion of the first neighbor hopping term,
\begin{equation}
V_1 = \Delta_1 \sin\frac{k_x}{2} \sin\frac{k_y}{2} \tau_x
\end{equation}
The Hamiltonian $H + V_1$ retains inversion $P$, along with $\{M_{\hat z} \vert\frac{1}{2} \frac{1}{2}\} $ and $\{M_{\hat x+\hat y}|00\}$, but the $C_2$ screw symmetries are violated.   As shown in Fig.~\ref{fig:bands_1}, the corner Dirac point is gapped, but the Dirac points at $X_{1,2}$ remain.  Provided there are no other extraneous electron and hole pockets and the electron count is $4n+2$, a system with these symmetries will be a 2D Dirac semimetal, with two symmetry equivalent Dirac points at $E_F$.   In fact, this is the {\it only} truly protected Dirac semimetal in the absence of approximate particle-hole symmetry, $t_2 \sim 0$.
If $M_{\hat x+\hat y}$ is violated, then the $X_{1,2}$ Dirac points are inequivalent, which belongs in the next case.

{\it Case II:  Two symmetry inequivalent Dirac points}.--We next consider breaking the glide mirror plane $\{M_{\hat z} \vert\frac{1}{2} \frac{1}{2}\} $  while keeping $\{C_{2\hat x}|\frac{1}{2}0\}$.  We displace the $B$ site in the $\hat y$ direction~[Fig.~\ref{fig:bands_2}], allowing a term
\begin{equation}
V_2 = \Delta_2 \cos\frac{k_x}{2}\sin\frac{k_y}{2} \tau_y.
\end{equation}
The Hamiltonian $H + V_2$ now has a gap at $X_2$, but Dirac points remain protected at $X_1$ and $M$, though in the presence of $t_2$ they are at different energies.  It follows from the arguments illustrated in Fig.~\ref{fig:cartoon} that a glide plane with fractional translation$\left(\frac{1}{2} 0\right)$ will produce this result as well. 

{\it Case III: Three Dirac points}.--All three Dirac points remain protected in the presence of $\{M_{\hat z} \vert\frac{1}{2} \frac{1}{2}\} $,
$\{C_{2\hat x}|\frac{1}{2}0\}$, and $\{C_{2\hat y}|0\frac{1}{2}\}$.   We find that the three Dirac points can persist even when the screw symmetries are violated provided the system retains an additional $C_{4\hat z}$ symmetry about the center of a plaquette.   This is violated by the crinkling responsible for $t^{\rm SO}$.  However, for the flat system $t_{\rm SO}=0$ this symmetry will pertain if the $A$ and $B$ sites have a lower symmetry internal structure
allowing a spin-orbit term 
\begin{equation}
V_3=t'_{\rm SO} \tau_z( -\sin k_x \sigma_x + \sin k_y \sigma_y)
\end{equation}
that preserves the three Dirac points.   Note, however, that it is impossible to gap the $X_1$ and $X_2$ points without also gapping $M$.   As discussed below, a single symmetry protected Dirac point is not possible.

These three cases represent all possible 2D systems with Dirac points: (1) symmetric Dirac points at both edges protected by $P$ and a $M_{\hat z}$ glide plane, (2) Dirac points at an edge and corner protected by $P$ and glide planes or screw axes along the $\hat{x}$ \emph{or} $\hat{y}$ directions, and (3) Dirac points at both edges and the corner protected by $P$ and a $C_{4\hat z}$ rotation, or case II for both the $\hat{x}$ \emph{and} $\hat{y}$ directions.   

\begin{figure}
\subfigure[]{\includegraphics[height=3.5cm]{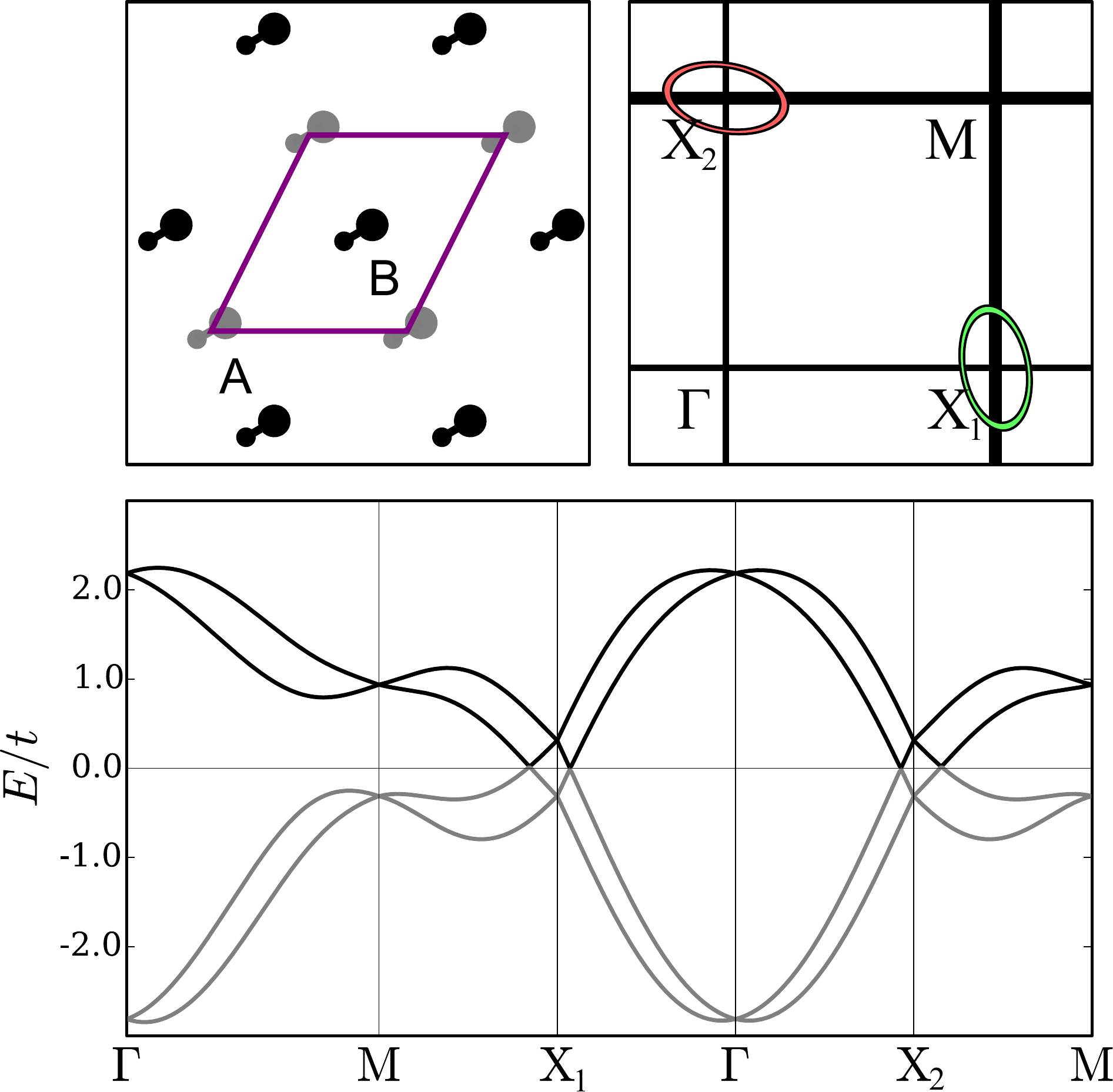}\label{fig:bands_M11}}
\subfigure[]{\includegraphics[height=3.5cm]{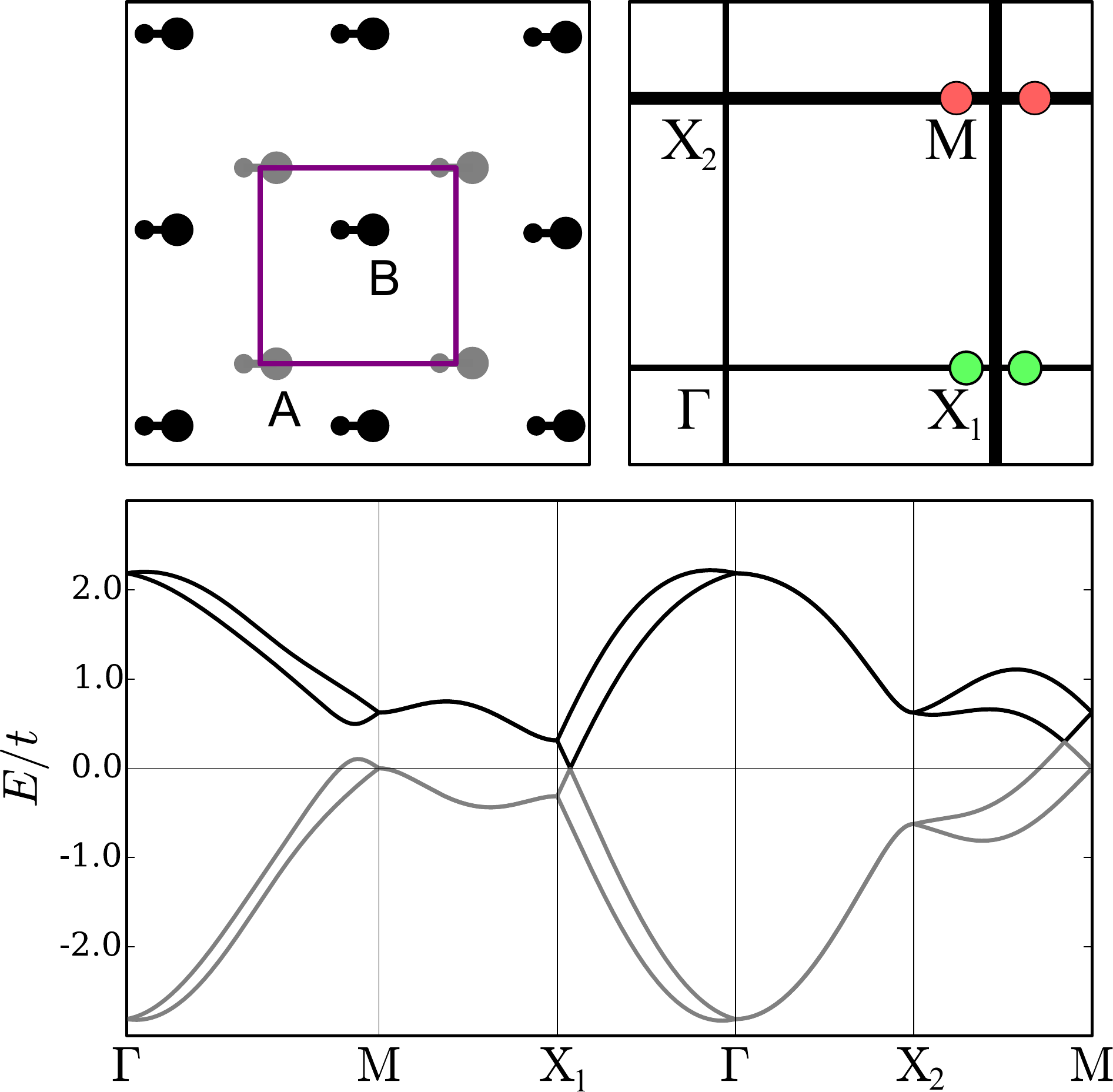}\label{fig:bands_M2}}

\caption{Without inversion, nonsymmorphic symmetries protect nodal lines or Weyl points.  This may be achieved by altering each site to be bipartite and asymmetric (e.g., heterodimers).~\subref{fig:bands_M11} Breaking inversion in case I, while preserving $\{M_{\hat z}|\frac{1}{2}\frac{1}{2}\}$, leads to nodal lines that circle the former Dirac points at $X_{1,2}$.~\subref{fig:bands_M2}  Breaking inversion in case II, preserving $\{C_{2\hat x}|\frac{1}{2}0\}$, results in Weyl points on the $C_{2\hat x}$ invariant lines.}
\end{figure}

{\it Case IV:  Line nodes and Weyl points}.--Finally, we mention that if inversion symmetry is violated, while keeping the nonsymmorphic symmetries, then the Dirac points are removed, but there remain Weyl points or line nodes.   For example, if in case I each site has a dipole moment $p (\hat x+\hat y)$, then a spin-orbit term $v_{\rm SO} \tau_x \sigma_z \sin(k_x-k_y)/2$ is allowed.   This preserves $\{M_{\hat z}|\frac{1}{2}\frac{1}{2}\}$ and for small $v_{\rm SO}$ leads to a circular line node surrounding the erstwhile Dirac points at $X_{1,2}$~[Fig.~\ref{fig:bands_M11}].    More generally for this symmetry, a line node will separate regions containing $\Gamma$ and $M$ from regions containing $X_{1,2}$.   Similarly, if $\{C_{2\hat x}|\frac{1}{2}0\}$ is preserved but not $\{M_{\hat z}|\frac{1}{2}\frac{1}{2}\}$, then the two Dirac points each split into two Weyl points along the $C_{2\hat x}$ invariant lines $k_y=0$ and $k_y=\pi$~[Fig.~\ref{fig:bands_M2}].
Weyl points at generic ${\bf k}$ are also locally protected when $\Theta$ and a $C_{2\hat z}$ symmetry are preserved even in the absence of nonsymmorphic symmetries.   However, in this case the Weyl points can annihilate, and are not guaranteed by symmetry.

\begin{figure}
\subfigure[]{\includegraphics[bb=-10 -20 100 65,width=4cm]{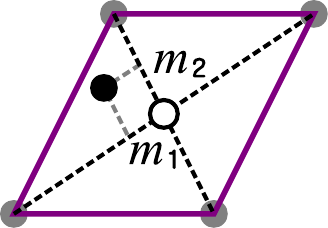}\label{fig:distort}}
\subfigure[]{\includegraphics[width=4cm]{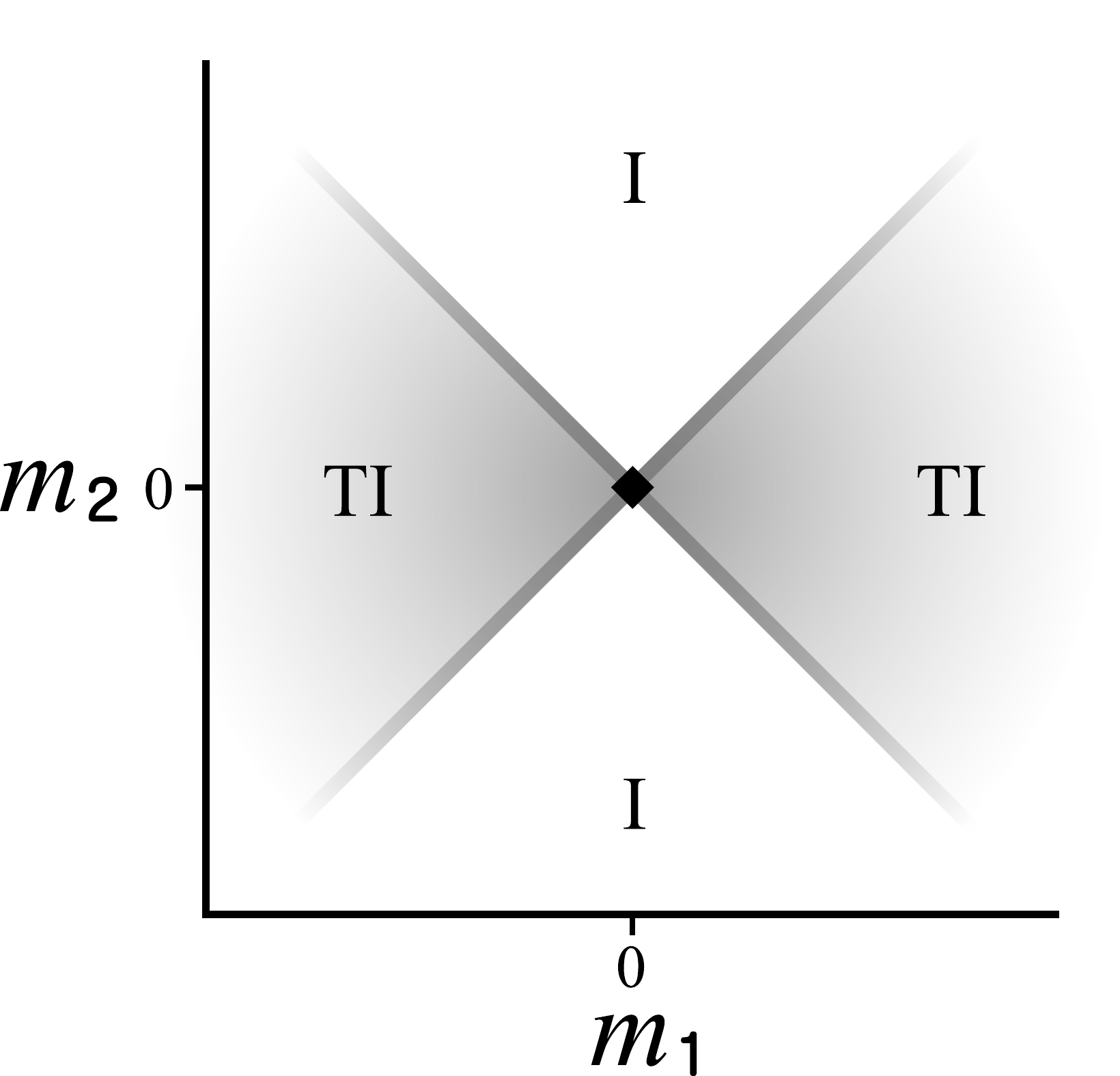}\label{fig:phase}}

\caption{By further lowering the symmetry the Dirac semimetal in Fig.~\ref{fig:bands_1} can be driven into either trivial (I) or topological (TI) insulator phases.   Lattice displacements in~\subref{fig:distort} give rise to the phase diagram in~\subref{fig:phase}.  }
\end{figure}

A distinctive feature of the nonsymmorphic Dirac semimetals is that they describe a critical point separating topologically distinct phases.  By lowering the symmetry it is possible to open a gap that leads to either a trivial or topological 2D insulator.  
Consider the system of case I~[Fig.~\ref{fig:bands_1}] with a gap introduced by displacing the center atom~[Fig.~\ref{fig:distort}], described by a perturbation
\begin{flalign*}
V_4 = \left[m_1\sin\left(\frac{k_x+k_y}{2}\right)+m_2\sin\left(\frac{k_x-k_y}{2}\right)\right]\tau_y
\end{flalign*}
where $m_1$ and $m_2$ describe the displacement in the  $[11]$ and $[\bar{1}1]$ directions, respectively. 
When $\abs{m_1}>\abs{m_2}$, the system is a topological insulator, and for $\abs{m_1}<\abs{m_2}$ the system is in a trivial phase~[Fig.~\ref{fig:phase}].  The boundaries between topological and trivial insulating phases are marked by a nonsymmetry-protected Dirac point.  Changing the sign of $m_1$ or $m_2$ results in a phase with the same topological character. However, one may gap the Dirac semimetal directly into either a topological or trivial insulator, depending on the direction of the displacement.  

It is worthwhile to contrast the 2D Dirac semimetal presented here with the symmetry protected Dirac points that arise at the surface of a weak topological insulator\cite{fukanemele} or a topological crystalline insulator\cite{futci}.  There, the surface Dirac points are protected by the combination of time-reversal and a translation or mirror symmetry.  Breaking the symmetry leads to topologically distinct gapped surface phases that map to each other under the symmetry.   Importantly, this leads to an absence of localization when the surface is strongly disordered, but retains the symmetry on the average\cite{stern, mong, qi,fukane2012}.   This phenomenon can occur only at the surface of a 3D topological phase.  In a purely 2D system the trivial and TI phases are {\it not} related by symmetry, so our 2D Dirac semimetal can be localized by strong disorder.   However, since it is in the symplectic
class, weak disorder leads to antilocalization. The absence of
symmetry relating the trivial and TI phases rules out a single symmetry protected Dirac point, since in that case the symmetry would change the sign of the single mass term.   We find that changing the sign of the mass of one of our Dirac points by a symmetry breaking perturbation always leads to the change in sign of another Dirac point, resulting in the same topological order.
Weak electron-electron interactions do not significantly alter the electronic structure of a 2D Dirac semimetal, though strong interactions could lead to a gapped state.  However, if the symmetry is not lowered that state must exhibit a nontrivial topological order\cite{ashvin}.

In terms of realistic materials, there are both challenges and advantages to working in 2D.  Two-dimensional systems can be more fragile, and typically require substrates that can influence the behavior.  On the other hand, they offer additional tunability not available in 3D systems.  For example, the proximity of TI and I phases could allow for flexible patterning of helical edge channels.   There has been surge of interest in monolayers of covalently bound layered materials over the past few years~\cite{Mak_2010}: these materials exhibit abundant variety in composition and structure~\cite{Ding_2011}, and have already shown hints of being able to host graphenelike Dirac points~\cite{Li_2014} and topological phases~\cite{Luo_2015,Qian_2014}. Many of these materials exist in structures belonging to appropriate symmetry groups, including litharge (case III) and WTe$_2$~\cite{Lee_2015} (case II), and some are known to possess the appropriate band filling, such as (Nb,Ta)Te$_2$~\cite{Brown_1966}. 
Moreover, the Dirac points in iridium oxide superlattices proposed by Chen and Kee\cite{kee} constitute a manifestation of case I described above.  They showed that distortions can lead to a TI gapped phase.   It will also be interesting to demonstrate the distortions in that system that lead to the trivial insulator.
We are thus optimistic about the prospects for the experimental study of 2D Dirac semimetals.

\begin{acknowledgements}
We thank Saad Zaheer for emphasizing to us the role of nonsymmorphic symmetries in Dirac semimetals.   We also thank Youngkuk Kim for helpful discussions.   C.L.K. acknowledges a Simons Investigator grant from the Simons Foundation, and S.M.Y. was supported by a National Research Council Research Associateship Award at the U.S. Naval Research Laboratory.
\end{acknowledgements}

%

\end{document}